\documentclass[arps,twocolumn,showpacs]{revtex4}
\input{epsf}
\usepackage{subfigure}

\newcommand\lsim{\mathrel{\rlap{\lower4pt\hbox{\hskip1pt$\sim$}}
        \raise1pt\hbox{$<$}}}
\newcommand\gsim{\mathrel{\rlap{\lower4pt\hbox{\hskip1pt$\sim$}}
        \raise1pt\hbox{$>$}}}
\def\ba{\begin{eqnarray}}
\def\ea{\end{eqnarray}}
\def\be{\begin{equation}}
\def\ee{\end{equation}}
\def\nn{\nonumber\\}
\def\thetaB{\mbox{\boldmath$\theta$}}
\def\Cgg{C^{gg}_\ell}
\def\Cnn{C^{nn}_\ell}
\def\Cgm{C^{g\mu}_\ell}
\def\Cmm{C^{\mu\mu}_\ell}

\begin{document}
\title{Lensing corrections to features in the angular two-point correlation function and power spectrum}
\author{Marilena LoVerde$^{1,2}$, Lam Hui$^{1,2,3}$, Enrique Gazta\~{n}aga$^{4}$}

\affiliation{
$^{1}$Institute for Strings, Cosmology and Astro-particle Physics (ISCAP)\\
$^{2}$Department of Physics, Columbia University, New York, NY 10027\\
$^{3}$Institute of Theoretical Physics, The Chinese University of Hong Kong\\
$^{4}$Institut de Ci\`encies de l'Espai, CSIC/IEEC, Campus UAB,
F. de Ci\`encies, Torre C5 par-2,  Barcelona 08193, Spain\\
marilena@phys.columbia.edu, lhui@astro.columbia.edu, gazta@aliga.ieec.uab.es
}
\begin{abstract}
It is well known that magnification bias, the modulation of galaxy or quasar source counts by gravitational lensing, can change the observed angular correlation function. We investigate magnification-induced changes to the shape of the observed correlation function $w(\theta)$, and the angular power spectrum $C_{\ell}$, paying special attention to the matter-radiation equality peak and the baryon wiggles. 
Lensing effectively mixes the correlation function of the source galaxies with that of the
matter correlation at the lower redshifts of the lenses distorting the observed correlation function. We quantify how the lensing corrections depend on the width of the selection function, the galaxy bias $b$, and the number count slope  $s$.  The lensing correction increases with redshift and larger corrections are present for sources with steep number count slopes and/or broad redshift distributions. The most drastic changes to  $C_{\ell}$ occur for measurements at high redshifts ($z\gsim1.5$) and low multipole moment ($\ell \lsim 100$). For the source distributions we consider, magnification bias can shift the location of the matter-radiation equality scale by $1$--$6\%$ at $z\sim1.5$ and by $z\sim3.5$ the shift can be as large as $30\%$. The baryon bump in $\theta^2w(\theta)$ is shifted by $\lsim 1\%$ and the width is typically increased by $\sim10\%$.  Shifts of $\gsim 0.5\%$ and broadening $\gsim 20\%$ occur only for very broad selection functions and/or galaxies with $(5s-2)/b\gsim2$.  However, near the baryon bump the magnification correction is not constant but is a gently varying function which depends on the source population. Depending on how the $w(\theta)$ data is fitted, this correction may need to be accounted for when using the baryon acoustic scale for precision cosmology. 

\end{abstract}
\maketitle
\section{Introduction}
The galaxy two-point angular correlation function and its spherical harmonic transform, the angular power spectrum, provide information about dark matter clustering. 
These statistics have scale dependent features which may be used as ``standard rulers'' to measure cosmological distances  \cite{Peebles1973,Cooray06}. Features in the three-dimensional power spectrum which appear at a comoving wave number $k$ will appear in the angular power spectrum at redshift $z_0$ at multipole $\ell \sim k\chi(z_0)$, where $\chi(z_0)$ is the comoving distance to $z_0$. One such feature is the peak in the power spectrum that separates the modes which entered the horizon during radiation domination
from those modes that entered during matter domination. This scale is determined by the horizon size at matter-radiation equality. Also present in the power spectrum are the so-called baryon acoustic oscillations (BAO). This series of wiggles 
in Fourier space is a signature of acoustic oscillations in the photon-baryon fluid that was present in the early universe.  The location of the peaks of the BAO depends on the horizon size at the time of recombination \cite{BAO1,BAO2,BAO3,BAO4,BAO5,EH98,BAO6}. This scale is robustly measured from the cosmic microwave background \cite{WMAPI,WMAPIII}, and thus can serve as a ``standard ruler''; given the scale of the BAO a measurement of the angular size of the BAO at some redshift will yield the angular diameter distance to that redshift . In fact the first measurements of this peak have occurred recently \cite{BAOmeas1,BAOmeas2,BAOmeas3,BAOmeas4,BAOmeas5}. In recent years, much effort has gone towards using these features in the correlation function for precision measurements \cite{BAOth1}.

Gravitational lensing changes the observed number density of galaxy or quasar sources - an effect called magnification bias  \cite{Gunn67,Narayan1989,BTP1996}. (Hereafter, the terms `galaxy' and `quasar' can be considered synonymous.)
It is well known that magnification bias modifies the galaxy angular correlation function \cite{Villumsen1995,VFC97,MJV98,MJ98,EGmag03,ScrantonSDSS05,menard,JSS03}. 
In this paper we extend the previous analyses by investigating and quantifying how magnification bias changes the shape of the angular two point correlation function, and its spherical harmonic transform the angular power spectrum, paying
special attention to important features such as the turnover in the power spectrum and the BAO.

Corrections from gravitational lensing enter as follows. The two-point function is measured from the galaxy number density fluctuation. 
\be
\delta_n=\frac{n({\bf x},z)-\bar{n}(z)}{\bar{n}(z)}
\ee
The effect of gravitational lensing is to alter the area of the patch of sky being observed and to change the observed flux of the source. Both effects change the measured galaxy number density, the first by changing the area, the second by changing the number of sources observed in a flux limited survey. Together these effects are called magnification bias \cite{Gunn67,Narayan1989,Villumsen1995,BTP1996}. To first order
(i.e. the weak lensing limit), they lead to a correction term $\delta_\mu$ being added to the intrinsic galaxy fluctuation $\delta_g$
\be
\delta_n=\delta_g+\delta_\mu
\label{deltan}
\ee
With this term the observed autocorrelation function becomes, 
\ba
\langle\delta_n\delta_n\rangle=\langle\delta_g\delta_g\rangle+\langle\delta_g\delta_\mu\rangle+\langle\delta_\mu\delta_g\rangle+\langle\delta_\mu\delta_\mu\rangle.
\ea
The galaxy-galaxy term $\langle\delta_g\delta_g\rangle$ depends on the matter distribution at the source galaxies. The magnification terms,  especially $\langle\delta_\mu\delta_\mu\rangle$, depend on the matter distribution spanning the range between the sources and the observer. 

The lensing of high redshift quasars by low redshift galaxies have been detected confirming the presence of magnification bias for these systems. The most recent measurements of this effect are discussed
in \cite{EGmag03,JSS03,ScrantonSDSS05} (see also \cite{Myers2003}). Discussions of earlier measurements can be found in the references therein. These measurements work by cross-correlating the angular positions of galaxies/quasars
at widely separated redshifts. Here, we will focus on the angular correlation of
galaxies at similar redshifts. As was first pointed out by \cite{Matsubara}, magnification bias alters observations of the 3D clustering of galaxies.  In two separate papers \cite{3Dpaper1,3Dpaper2}, we further consider the effect of magnification bias on the 3D correlation function and power spectrum, which turn out to have qualitatively new and interesting features.

While this paper was in preparation, a paper addressing weak gravitational lensing corrections to baryon acoustic oscillations was posted \cite{otherpaper}. In that paper they consider the effect of magnification bias (a first order correction) and stochastic deflection (a second order correction) on BAO in the real space correlation function for a delta function source distribution. In contrast, we consider the effect of magnification bias on the angular correlation function and the angular power spectrum for an extended source distribution. As we will show the width of the source distribution strongly affects the magnitude of the magnification bias correction. While some conclusions we reach are addressed in \cite{otherpaper}, the observables considered and the analyses here are different. 

This paper is organized as follows. In section \ref{corrfunctions} we present expressions for the angular auto-correlation function and the angular power spectrum when magnification bias is included. In section \ref{ampsection} we discuss the factors affecting the relative magnitude of the magnification bias terms compared with the galaxy term. In section \ref{shapesection} we examine the effect of magnification on the shape of the angular power spectrum for a few different source distributions. In section \ref{BAOsec} we consider the effect of magnification on the baryon bump in the angular correlation function. In \ref{discussion} we conclude and discuss the implications of our work for future surveys. 

\section{Anisotropies and Correlation functions} 
\label{corrfunctions}
We consider the two-dimensional galaxy fluctuation in direction $\hat{\thetaB}$ integrated over a selection function with mean redshift $z_0$
\be
\label{deltandef}
\delta_n(\hat{\thetaB},z_0)=\frac{n(\hat{\thetaB},z_0)-\bar{n}(z_0)}{\bar{n}(z_0)}
\ee
where $\bar{n}$ is the mean number of galaxies in the redshift bin. We include this label, $z_0$, to remind the reader that the angular fluctuation depends on the source selection function. 

When the effect of magnification bias is considered, the expression for the net measured galaxy overdensity becomes a sum of two terms (Equation \ref{deltan}). In the expressions that follow and throughout this paper we assume a flat universe, generalizing to an open or closed universe is fairly straightforward. 
%\begin{equation}
%\label{deltan}
%\delta_n(\hat{\thetaB},z_0)=\delta_{g}(\hat{\thetaB},z_0)+\delta_{\mu}(\hat{\thetaB},z_0)
%\end{equation}
The first term in Equation \ref{deltan} is the intrinsic galaxy fluctuation integrated over a normalized selection function given by $W(z,z_0)$,
\begin{equation}
\label{deltag}
\delta_g(\hat{\thetaB},z_0)= \int_{0}^\infty\!\! dz \, b(z) W(z,z_0) \delta(\chi(z)\hat{\thetaB},z).
\end{equation}
Where $\chi(z)$ is the comoving distance to redshift $z$, $\delta=(\rho-\bar{\rho})/\bar{\rho}$ is the matter overdensity and $b(z)$ is a bias factor relating the galaxy number density fluctuation to the matter density fluctuation $\delta_g=b(z)\delta$. For simplicity we have assumed that the bias is scale independent, this should be accurate for $k\lsim 0.05 h\textrm{Mpc}^{-1}$, however at smaller scales it may be important (see for example \cite{RS3}). If we further assume that $b(z)$ is slowly varying across $W(z,z_0)$ then we can set $b(z)=b(z_0)$ and pull it out of the integral in Equation \ref{deltag}. 
%\begin{equation}
%\delta_g(\hat{\thetaB},z_0)= b(z_0)\int_{0}^\infty\!\! dz \, W(z,z_0) \delta(\chi(z)\hat{\thetaB},z).
%\end{equation}

The second term in Equation \ref{deltan} is the correction due to magnification bias,
\begin{eqnarray}
\label{mutermfull}
\delta_{\mu}(\hat{\thetaB},z_0)=\int_0^{\infty}\!\!dz\frac{c}{H(z)}\nabla_{\perp}^2\phi(\chi(z)\hat{\thetaB},z_0)\qquad\qquad\qquad\\ \nonumber
\times\,\,\chi(z)\int_z^{\infty} dz'\left(5s(z')-2\right)\frac{\chi(z')-\chi(z)}{\chi(z')}W(z',z_0). 
\end{eqnarray}
$H(z)$ is the Hubble parameter and $c$ is the speed of light \cite{Narayan1989}. The magnification bias term depends on the Laplacian of the gravitational potential $\phi$ (with respect to the comoving coordinates in the direction perpendicular to $\hat{\thetaB}$) and the slope of the number count function. For a survey with limiting magnitude $m$ this is
\begin{equation}
s=\frac{d\textrm{log}_{10}N(<m)}{dm}.
\end{equation}
If $s(z)$ is slowly varying across $W(z,z_0)$, then we can set $s(z)=s(z_0)$ and the expression for $\delta_\mu$ becomes
\begin{eqnarray}
\label{muterm}
\delta_{\mu}(\hat{\thetaB},z_0)&=&\left(5s(z_0)-2\right)\nonumber\\
&\times&\int_0^{\infty}\!\!dz\frac{c}{H(z)}g(z,z_0)\nabla_{\perp}^2\phi(\chi(z)\hat{\thetaB},z_0).
\end{eqnarray}
Where we have introduced the lensing weight function
\begin{equation}
\label{lenswt}
g(z,z_0)=\chi(z)\int_z^{\infty} dz'\frac{\chi(z')-\chi(z)}{\chi(z')}W(z',z_0). 
\end{equation}
The lensing weight function can be thought of as roughly proportional to 
the probability for sources in $W(z,z_0)$ to be lensed by density perturbations at $z$. The lensing weight function increases in magnitude as $z_0$ increases and is peaked at a redshift $z$ corresponding to about half the comoving distance to $z_0$. 

On scales $k\gg aH$ Poisson's equation can be used to relate the gravitational potential to the matter fluctuation
\be
-k^2 \phi({\bf k},z)=\frac{3H_0^2}{2c^2} \Omega_m (1+z)\delta({\bf k},z)
\ee
where ${\bf k}$ is the comoving wave vector, $k$ is its magnitude and 
$\Omega_m$ is the matter density today \cite{dodelson}. 
.

We consider the two-point correlation function of the galaxy overdensity 
\begin{eqnarray}
\label{wtheta1}
w_{nn}(\theta,z_0)&=&\langle\delta_g(\hat{\thetaB},z_0)\delta_g(\hat{\thetaB}',z_0)\rangle+2\langle\delta_g(\hat{\thetaB},z_0)
\delta_{\mu}(\hat{\thetaB}',z_0)\rangle\nonumber\\
&+&\langle\delta_{\mu}(\hat{\thetaB},z_0)\delta_{\mu}(\hat{\thetaB}',z_0)\rangle\nonumber\\
&\equiv& w_{gg}(\theta,z_0)+2w_{g\mu}(\theta,z_0)+w_{\mu\mu}(\theta,z_0)
\end{eqnarray}
where $\cos\theta=\hat{\thetaB}\cdot\hat{\thetaB}'$.   In what follows we will consider the Legendre coefficients of the correlation functions. These are defined as
\be
\label{wthetaeq}
w_{gg}(\theta,z_0)=\sum_{\ell}\frac{2\ell+1}{4\pi} C^{gg}_{\ell}(z_0)P_{\ell}(\cos\theta)
\ee
where $P_\ell(\cos\theta)$ are the Legendre polynomials. The Legendre components of the observed correlation function, $w_{nn}(\theta,z_0)$ will of course be a sum of all the terms,
\be
C^{nn}_{\ell}(z_0)=C^{gg}_{\ell}(z_0)+2C^{g\mu}_{\ell}(z_0)+C^{\mu\mu}_{\ell}(z_0)
\ee
We calculate the angular power spectra, $\Cgg$, $\Cgm$ and $\Cmm$, using the Limber approximation \cite{Limber} which is accurate for $\ell\gsim 10$. To simplify the expressions and provide some insight into the effects of the magnification terms we introduce the following functions
%\begin{eqnarray}
%\label{phiggeq}
%\phi^{gg}(z,z_0)=b^2W(z,z_0)^2\,\qquad\qquad\qquad\qquad\qquad\qquad\quad\\
%\label{phigmeq}
%\phi^{g\mu}(z,z_0)=\qquad\qquad\qquad\qquad\qquad\qquad\qquad\qquad\qquad\,\\
%\frac{3}{2}b(5s-2)\Omega_m\frac{H_0^2}{c H(z)}W(z,z_0)g(z,z_0)(1+z)\nn
%\label{phimmeq}
%\phi^{\mu\mu}(z,z_0)=\qquad\qquad\qquad\qquad\qquad\qquad\qquad\qquad\qquad\,\\
 %\frac{9}{4}(5s-2)^2\Omega^2_m \frac{H_0^4}{c^2H(z)^2}g^2(z,z_0)(1+z)^2\nn
%\end{eqnarray}

\begin{eqnarray}
\label{phiggeq}
\phi^{gg}(z,z_0)=b(z_0)^2W(z,z_0)^2\,\qquad\qquad\qquad\qquad\qquad\quad\,\,\\
\label{phigmeq}
\label{phimmeq}
\phi^{\mu\mu}(z,z_0)=\qquad\qquad\qquad\qquad\qquad\qquad\qquad\qquad\qquad\,\\
 \frac{9}{4}(5s(z_0)-2)^2\Omega^2_m \frac{H_0^4}{c^2H(z)^2}g^2(z,z_0)(1+z)^2\nn
 \phi^{g\mu}(z,z_0)=\sqrt{\phi^{gg}(z,z_0)\phi^{\mu\mu}(z,z_0)}\qquad\qquad\qquad\qquad\quad
\end{eqnarray}

With this notation the angular correlation functions can all be expressed as
\be
\label{cxxleq}
C^{xx}_{\ell}(z_0)=\int_0^\infty \frac{dz}{\chi(z)^2}\frac{H(z)}{c}\phi^{xx}(z,z_0)P\left(\frac{\ell}{\chi(z)},z\right)
\ee
where $xx$ symbolizes $gg$, $g\mu$ or $\mu\mu$ and $P(k=\ell/\chi(z),z)$ is the matter power spectrum. 

\begin{figure}[tb]
\centerline{\epsfxsize=9cm\epsffile{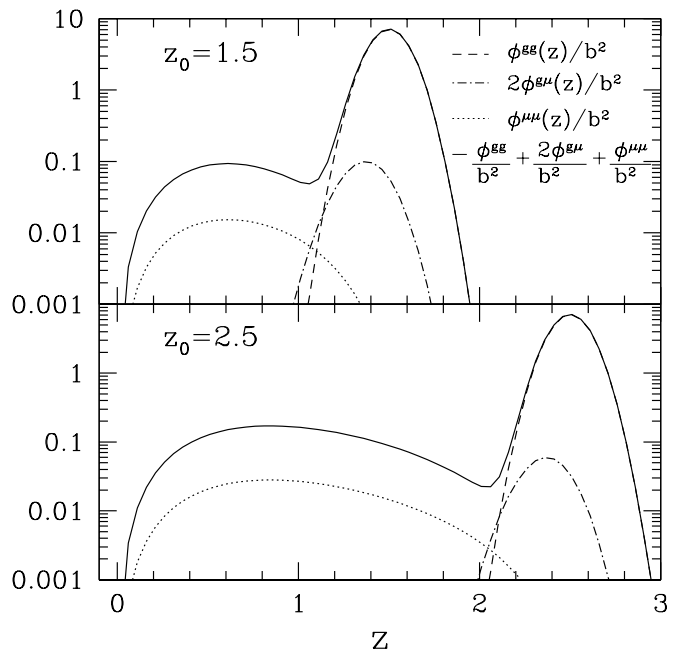}}
\caption{The redshift distributions normalized by the galaxy bias (Eq. \ref{phiggeq}-\ref{phimmeq}), $\phi^{gg}/b^2$ (dashed line), $2\phi^{g\mu}/b^2$ (dot-dashed line) and $\phi^{\mu\mu}/b^2$ (dotted line) shown here for Gaussian selection functions centered at $z_0=1.5$ and $z_0=2.5$ with $\sigma=0.15$. These distributions determine the angular power spectra through Equation \ref{cxxleq}. The total auto-correlation $C^{nn}_\ell/b^2$ will depend on $(\phi^{gg}+2\phi^{g\mu}+\phi^{\mu\mu})/b^2$ (solid line). }
\label{SelRad}
\end{figure}

For illustration we use Gaussian selection functions
\be
\label{sel}
W(z,z_0)=\frac{1}{\sqrt{2\pi}\sigma}\textrm{exp}\left[-\frac{(z-z_0)^2}{2 \sigma^2}\right]
\ee
of various widths to demonstrate how the shape of the selection function alters the effect of magnification.

We assume a flat $\Lambda$CDM cosmology with $\Omega_m=0.27$, $\Omega_{\Lambda}=0.73$ and $\Omega_b=0.0224/h^2$ as the fractional energy densities in matter, vacuum and baryons today.  The Hubble constant $H_0=100h$ km/s/Mpc 
is set to $h=0.7$, the fluctuation amplitude to $\sigma_8=0.8$, and scalar spectral index $n_s=0.95$. For the linear power spectrum we use the Eisenstein and Hu transfer function \cite{EH98}. In a few places (e.g. Figure \ref{CnnCnoBfig}) we discuss the no-BAO spectrum, this is calculated using the BBKS transfer function \cite{BBKS} with a modified shape parameter $\Gamma=\Omega_mh\,\textrm{exp}[-\Omega_b(1+\sqrt{2h}/\Omega_m)]$ \cite{GammaSugiyama}. The non-linear evolution of both power spectra is calculated using the prescription of Smith \emph{et. al.} \cite{Smithetal}. The nonlinear power spectrum is used in all plots and discussions.

Unless otherwise stated we set the galaxy sample dependent ratio $(5s-2)/b=1$ (see \cite{mepaper,3Dpaper1} on how this varies with galaxy sample and redshift). The first correction term, $\Cgm/b^2$ is linear in this quantity while the second term, $\Cmm/b^2$ is quadratic.  Thus the magnitude of the net correction,  calculated here cannot in general be scaled by $(5s-2)/b$.  However, at low redshifts $2\Cgm/b^2$ dominates over $\Cmm/b^2$ and at higher redshifts $\Cmm > 2\Cgm$. The redshift of the transition between the two regimes increases with the width of the selection function. Specifically when $\sigma=0.07$, $2\Cgm <\Cmm$ for $z_0 \ge 1.5$, when $\sigma=0.15$, $2\Cgm <\Cmm$ for $z_0\ge 2.0$ and when $\sigma=0.30$, $2\Cgm <\Cmm$ for  $z_0\ge 2.5$.

\section{amplitude of the lensing corrections}
\label{ampsection}
The relative magnitude of the lensing magnification terms, $\Cgm$ and $\Cmm$, to the intrinsic galaxy term, $\Cgg$, depends on several things: first the galaxy-sample dependent quantities $b$ and $s$, and
second the selection function and cosmological quantities in Equations \ref{phiggeq}--\ref{phimmeq}. Here we discuss how these quantities affect the relative magnitudes of $\Cgg$, $\Cgm$ and $\Cmm$.

The magnification terms are scaled by the galaxy-sample dependent factors $(5s-2)/b$ and $(5s-2)^2/b^2$. Thus for a given galaxy bias, galaxies residing on the steep end of the luminosity function will have larger magnification corrections. However, if $s<2/5$, the galaxy-magnification term $\Cgm$ is negative, while the magnification-magnification term, $\Cmm$ is always positive. Unless otherwise stated we fix $(5s-2)/b=1$ so both terms are positive. 

The difference between the galaxy-galaxy, galaxy-magnification and magnification-magnification terms can be seen more clearly by considering equations \ref{phiggeq}--\ref{cxxleq}. From Equation \ref{cxxleq}, we see that all the angular power spectra are simply integrals of the power spectrum over some redshift distribution $\phi^{xx}(z,z_0)$. Examples of these distributions for a selection function with $\sigma=0.15$ are shown in Figure \ref{SelRad}. In the case of $\phi^{gg}(z,z_0)$ this distribution is determined by the selection function. This suggests that one can think of the magnification terms as adding to $\Cgg$ measurements of the correlation function with ``selection functions'' determined by $\phi^{g\mu}(z,z_0)$ and $\phi^{\mu\mu}(z,z_0)$. 

Figure \ref{SelRad} shows that $\phi^{g\mu}(z,z_0)$ is peaked not too far from where the selection function is peaked, and is similar in shape to the selection function. Thus we expect $\Cgm$ to be similar in shape to $\Cgg$. On the other hand, the magnification-magnification term $\phi^{\mu\mu}$ is peaked at a much lower redshift than the selection function and is also much more broadly distributed in redshift. Thus we expect $\Cmm$ to be peaked at lower $\ell$ (larger angular scales) than $\Cgg$ because it is probing structure at lower redshifts which is nearer to the observer and occupies a larger angle in the sky. Additionally, since $\phi^{\mu\mu}(z,z_0)$ is rather broadly distributed, we expect sharp features such as baryon oscillations to be smoothed out 
in $\Cmm$.

One might infer from looking at Figure \ref{SelRad} that the magnification bias terms $\Cgm$ and $\Cmm$ will be quite small compared to $\Cgg$. However, the other quantities in the integrand of Equation \ref{cxxleq} decrease with redshift so the relative magnitudes of $\phi^{gg}$, $\phi^{g\mu}$ and $\phi^{\mu\mu}$ do not completely determine the relative magnitudes of $C^{gg}_{\ell}$,$C^{g\mu}_{\ell}$ and $C^{\mu\mu}_{\ell}$. 

\section{the shape of the Angular power spectrum}
\label{shapesection}

\begin{figure}[tb]
\centerline{\epsfxsize=9cm\epsffile{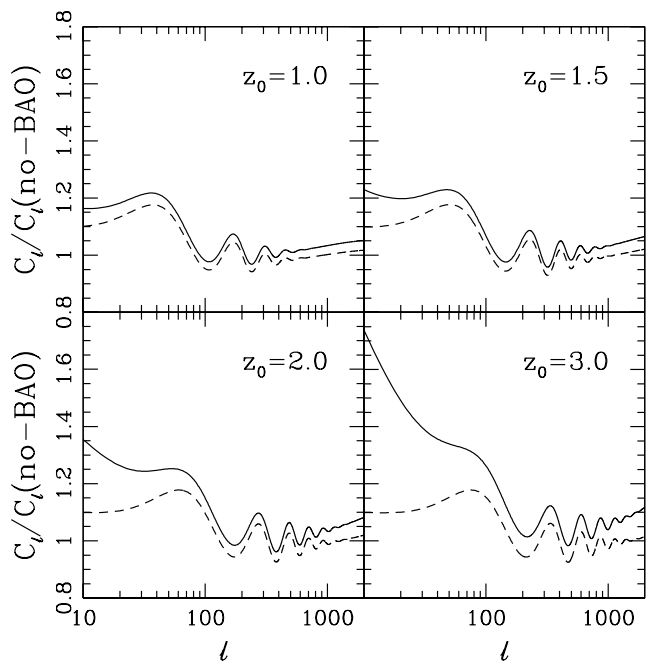}}
\caption{The angular power spectra divided by the no-BAO power spectra for a selection function of width $\sigma=0.15$. The dashed line is $\Cgg$ alone (no magnification case), the solid line is $\Cnn=\Cgg+2\Cgm+\Cmm$.}
\label{CnnCnoBfig}
\end{figure}

Using equations \ref{phiggeq}--\ref{phimmeq} and \ref{cxxleq} we calculate the angular power spectra using Gaussian selection functions centered at a variety of redshifts. Because the galaxy-magnification cross term $\Cgm$ depends strongly on the width of the selection function,
we consider several different widths given by $\sigma=0.07,0.15,0.30$. The magnification-magnification term $\Cmm$ is largely independent of the width of the selection function, but the galaxy-magnification term $\Cgm$ increases with increasing width of the selection function.
The effect of increasing $\sigma$ is therefore to increase the overall contribution of magnification to $C^{nn}_\ell$.

In Figure \ref{CnnCnoBfig} we show the angular power spectrum with and without magnification bias for a selection function of width $\sigma=0.15$. In this figure the power spectra are normalized by the no-BAO power spectrum, which is calculated neglecting magnification bias and the effects of baryons. We see immediately several features: the effect of magnification is largest at the low $\ell$ values (small $\ell$ correspond to large angular scales), the magnitude of the magnification correction increases with redshift, and the effect at high $\ell$ near the baryon wiggles is mostly (though not completely) to boost the amplitude rather than change the shape of the spectrum. 

Magnification clearly changes the shape of the angular power spectrum. The most significant changes are at $\ell\lsim 100$ where the peak in the power spectrum resides. The peak in the power spectrum is an important feature because its location is related to the Hubble scale when the universe transitions from radiation domination to matter domination.  This feature in the power spectrum can be used as a standard ruler \cite{Cooray06}. 

We identify the location $\ell_{peak}$ and amplitude of the peak in $\Cgg$ and $\Cnn$.
The shift in $\ell_{peak}$, and the fractional change in amplitude, $(\Cnn-\Cgg)/\Cgg$ (where $\Cnn$ is evaluated at the $\ell_{peak}$ for $\Cnn$ and $\Cgg$ is evaluated at the $\ell_{peak}$ found for $\Cgg$) when magnification is included are shown in Table \ref{lpeaktable}. The change in $\ell_{peak}$ and the change in amplitude increase both with redshift and with increasing width of the selection function $\sigma$.  The value of $\ell_{peak}$ is also dependent on the width of the selection function, with $\ell_{peak}$ taking slightly larger values for narrower selection functions.  %For the narrowest selection function with $\sigma=0.07$ the shift in $\ell_{peak}$ due to magnification remains $\lsim 7\%$ for all redshifts, but for the broadest selection function with $\sigma=0.30$ the shift is nearly $7\%$ by redshift $z_0=1.0$.

One additional consequence of magnification bias is to change the redshift and scale at which non-linear evolution of the power spectrum becomes important. This is because even for sources at high redshift where non-linearity is small, $\Cmm$ will depend on structure at low redshift where non-linearity is important. Indeed, even at $z_0=3.5$ the difference between $\Cnn$ calculated with the non-linear power spectrum and $\Cnn$ calculated with the linear power spectrum is $\gsim 10\%$ by $\ell \sim 1800, 1400,1000$ for $\sigma =0.07$, $0.15$ and $0.30$ respectively. This is to be compared with $\Cgg$, the no magnification case, for which the difference between $\Cgg(z_0=3.5)$ calculated with and without non-linear evolution does not approach $10\%$ until $\ell \sim 2600$ for each value of $\sigma$.

\begin{table*}
\begin{small}
\begin{tabular}{|c|c|c|c|c|c|c|c|c|c|}
\hline
&\multicolumn{3}{|c|}{$\sigma=0.07$} &\multicolumn{3}{|c|}{$\sigma=0.15$} & \multicolumn{3}{|c|}{$\sigma=0.30$} \\
\hline 
$z_0$ &	$\ell_{peak}$	& 	$\Delta\ell_{peak}$	&	$\Delta C_{\ell_{peak}}/C^{gg}_{\ell_{peak}}$	&	$\ell_{peak}$	&	$\Delta \ell_{peak}$	&	$\Delta C_{\ell_{peak}}/C^{gg}_{\ell_{peak}}$ &	$\ell_{peak}$	&	$\Delta \ell_{peak}$	&	$\Delta C_{\ell_{peak}}/C^{gg}_{\ell_{peak}}$ \\
\hline
$0.5$	&	$19$	&	$0$	&	$0.02$	&	$16$	&	$0$	&	$0.04$	&	$-$	&	$-$	&	$-$	\\
\hline
$1.0$	&	$34$	&	$0$	&	$0.01$	&	$33$	&	$0$	&	$0.04$	&	$30$	&	$-2$	&	$0.14$	\\
\hline
$1.5$	&	$46$	&	$-1$	&	$0.02$	&	$45$	&	$-1$	&	$0.05$	&	$44$	&	$-3$	&	$0.14$	\\
\hline
$2.0$	&	$55$	&	$-1$	&	$0.03$	&	$54$	&	$-3$	&	$0.07$	&	$54$	&	$-5$	&	$0.17$	\\
\hline
$2.5$	&	$62$	&	$-2	$	&	$0.04$	&	$61$	&	$-5$	&	$0.10$	&	$61$	&	$-9$	&	$0.22$	\\
\hline
$3.0$	&	$67$	&	$-3$	&	$0.06$	&	$67$	&	$-8$	&	$0.13$	&	$67$	&	$-15$	&	$0.29$	\\
\hline
$3.5$	&	$72$	&	$-5$	&	$0.08$	&	$72$	&	$-11$	&	$0.17$	&	$72$	&	$-21$	&	$0.38$	\\
\hline

\end{tabular}
\end{small}
\caption{A table of quantities related to the matter-radiation equality peak for a variety
of mean redshifts $z_0$ and selection function widths $\sigma$. 
The peak location (of $\Cgg$) is $\ell_{peak}$,
the shift in the peak location when magnification is included is
$\Delta \ell_{peak}=\ell_{peak}^{nn}-\ell_{peak}^{gg}$, and
the change in height at the peak is $\Delta C_{\ell_{peak}}/C_{\ell_{peak}}=\left(C^{nn}_{\ell_{peak}}-C^{gg}_{\ell_{peak}}\right)/C^{gg}_{\ell_{peak}}$.  For $z_0 = 0.5$, we do not consider the case of
a wide selection function $\sigma = 0.30$.}
\label{lpeaktable}
\end{table*}

\section{the angular correlation function and the baryon bump}
\label{BAOsec}

We now turn our attention to the real space angular correlation function $w_{nn}(\theta,z_0)$ (Equation \ref{wtheta1}). In multipole space, the baryon oscillations appear as a series of peaks in the power spectrum (e.g. Figure \ref{CnnCnoBfig}). In real space the signature of baryon oscillations is a single bump in the correlation function. The location of  the bump in $w_{nn}(\theta,z_0)$ is determined by the comoving sound horizon at recombination and the distance to $z_0$. The sound horizon at recombination is measured quite precisely from the cosmic microwave background anisotropy, thus a measurement of $\theta_{BAO}$ at $z_0$ can give a measure of the comoving distance to $z_0$. 

\begin{figure}
\centerline{\epsfxsize=9cm\epsffile{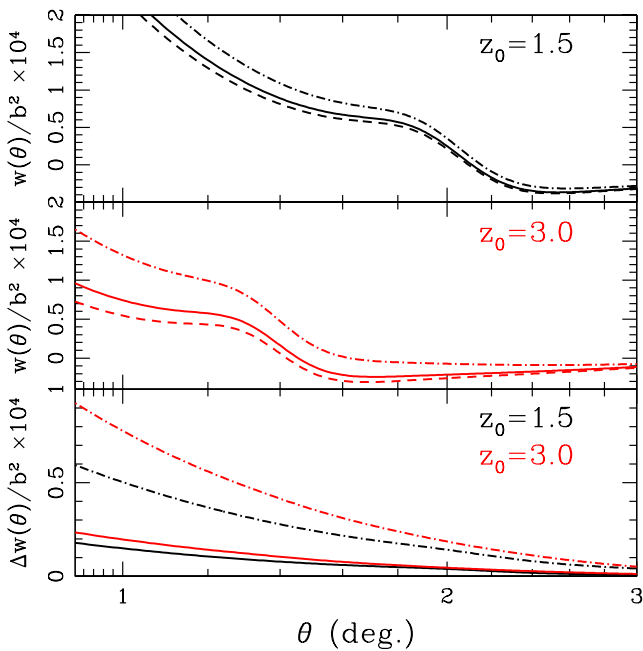}}
\caption{Top and middle panels: the two-point function divided by the galaxy bias $w(\theta)/b^2$.  This is using a selection function of width $\sigma=0.15$ centered at $z_0=1.5$ (top) and $_0=3.0$ (middle). The dashed line neglects magnification, the solid line includes magnification with $(5s-2)/b=1$, the dot-dashed line includes magnification with $(5s-2)/b=2$. Bottom panel: the magnification correction, $(w_{nn}(\theta)-w_{gg}(\theta))/b^2$. The solid line is for $(5s-2)/b=1$, the dot-dashed line is for $(5s-2)/b=2$}
\label{wggNear}
\end{figure}

\begin{figure}
\centerline{\epsfxsize=9cm\epsffile{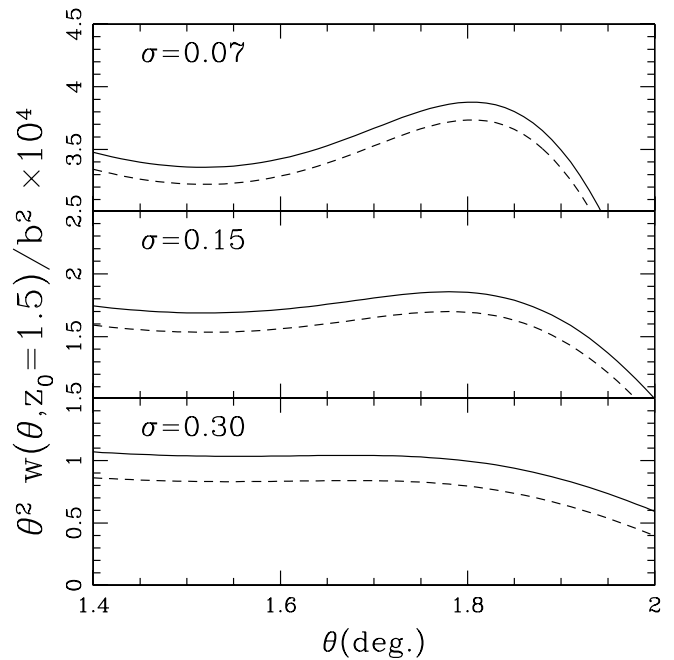}}
\caption{Here we zoom in on baryon bump in $\theta^2w(\theta)/b^2$ at $z_0=1.5$. The dashed lines neglect magnification $\theta^2w_{gg}(\theta)/b^2$ the solid lines include magnification $\theta^2w_{nn}(\theta)/b^2$. The broader the selection function, the harder it is to locate the baryon peak, the presence of magnification bias doesn't change this.}
\label{th2zoomfig}
\end{figure}

Before we address the lensing corrections to the baryon bump, we will first discuss a few issues that arise (whether or not lensing is included) when using the angular correlation function to measure the baryon oscillation scale. The angular correlation function is in some sense averaging the matter power spectrum across different redshifts. Thus, if too broad of a selection function is used the baryon feature will be washed out. For a fixed width in redshift, the washing out is more severe at low redshifts because $\Delta \chi \sim \Delta z /H(z)$ and $H(z)$ decreases with decreasing $z$. Additionally, the baryon bump is a subtle feature in the angular correlation function. 
It is customary to multiply the angular correlation function
by $\theta^2$ to make the baryon bump easier to identify and
characterize.
Indeed for the cases we have considered this is necessary in order for there to be a local maximum at the acoustic scale
at all. Figure \ref{wggNear} illustrates this point: while there is a baryon feature, there is no local maximum in $w(\theta)$. In Figure \ref{th2zoomfig} we show $\theta^2w(\theta)$ in the region near the baryon bump, in this case the baryon feature is quite visible for $\sigma=0.07$ but is very hard to identify for $\sigma=0.30$, fortunately the baryon bump becomes more prominent with increasing redshift. 

Since the baryon bump is broadened at low redshifts we limit our analysis to $z_0 \ge 1.0$ for $\sigma=0.07$ and $0.15$ and $z_0\ge 2.0$ for $\sigma=0.30$. We have experimented with multiplying by $\theta^3$: in this case even for the broadest selection function $\sigma=0.30$ the baryon bump is visible at all redshifts we consider. 
Interestingly, the magnitude of the shift in the baryon oscillation scale
due to magnification bias is sensitive to whether one looks at
$\theta^2 w(\theta)$ or $\theta^3 w(\theta)$. The shift is much larger in
the latter case, suggesting that the importance of 
magnification bias for baryon oscillation measurements depends
on precisely how the baryon bump is fitted.
In this paper, we focus on the effects on $\theta^2 w(\theta)$.

%\begin{figure}
%\centerline{\epsfxsize=9cm\epsffile{wgg.ps}}
%\caption{The two-point function divided by the galaxy bias and
%multiplied by angle: $\theta^2w(\theta)/b^2$. This is using a selection function of width $\sigma=0.15$ centered at $z_0=0.5$ for the upper panel and $z_0=1.5$ in the lower panel. The dashed line neglects magnification, the solid line includes magnification with $(5s-2)/b=1$.  For $z_0=0.5$, the baryon bump is difficult to identify.}
%\label{wgg}
%\end{figure}

We calculate the two-point function $w_{nn}(\theta,z_0)$ by performing the sum in Equation \ref{wthetaeq}.   In Figure \ref{wggNear} we plot $w_{nn}(\theta)$ and $w_{gg}(\theta)$ with redshift bins centered at $z_0=1.5$ and $z_0=3.0$ for a small angular range near the baryon bump. One can see from Figure \ref{wggNear} that, as expected, the angular correlation function is changed by magnification bias. 
%Figures \ref{th0wggdiff} and \ref{th2wggdiff} show the angular correlation function near the baryon bump and the magnification correction to the angular correlation function when $w(\theta)$ or $\theta^2w(\theta)$ is considered. 

The lensing correction to the angular correlation function is scale dependent. To address how the baryon bump is changed by lensing magnification we consider the location of the baryon peak $\theta_{BAO}$, the height at the peak $\theta^2w(\theta_{BAO})$, and the peak width. The peak width is defined to be $\left[\left(\frac{\partial^2}{d\theta^2} \theta^2w \right)/(\theta^2w)\right]^{-1/2}$ evaluated at $\theta_{BAO}$. 
\begin{figure}
\centerline{\epsfxsize=9cm\epsffile{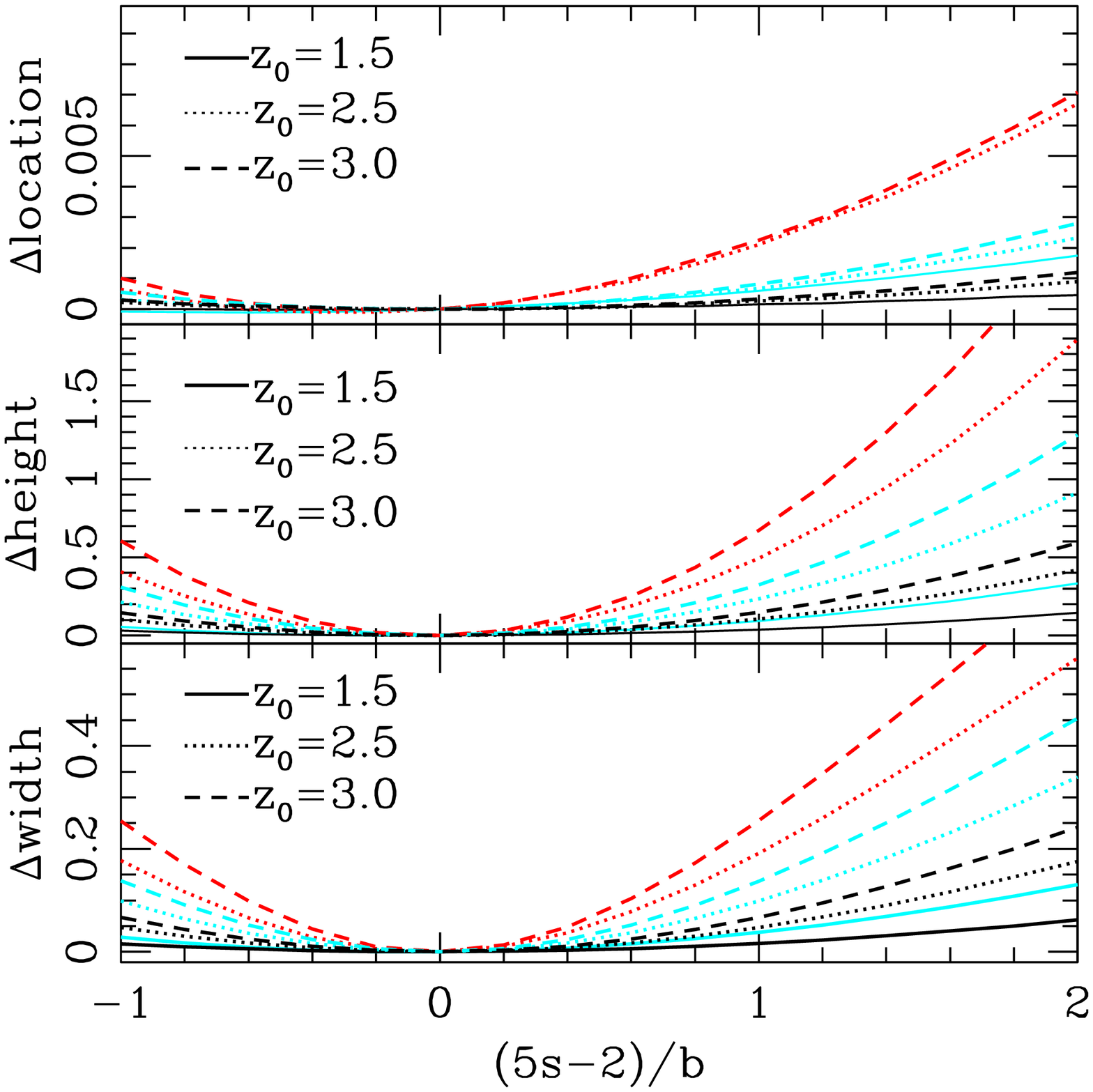}}
\caption{Effects of magnification bias on the baryon bump as found in $\theta^2w(\theta)$. Shown here for selection functions with $\sigma=0.07$ (black lowest curves), $\sigma=0.15$ (cyan middle curves), and $\sigma=0.30$ (red upper curves) as a function of the galaxy population-dependent ratio $(5s-2)/b$ for several different mean redshifts. Upper panel: the fractional change in the angular location of the baryon bump $(\theta_{BAO}^{nn}-\theta_{BAO}^{gg})/\theta_{BAO}^{gg}$. Middle panel: the fractional change in the height of the baryon bump, $(w_{nn}(\theta^{nn}_{BAO})-w_{gg}(\theta^{gg}_{BAO}))/w_{gg}(\theta_{BAO}^{gg})$. Bottom panel: the fractional change of the width of the baryon bump -- see text for how the width is defined.}
\label{th2DiffAll}
\end{figure}

Figure \ref{th2DiffAll} summarizes the fractional changes in the peak location, height and width as a function of $(5s-2)/b$ for selection functions with $\sigma=0.07$, $0.15$ and $0.30$,
and for several redshifts. The magnitude of the changes to the peak increase with redshift for all $\sigma$. The changes in peak location and width are largest at the highest redshifts ($z_0=3-3.5$) but the exact redshift dependence varies depending on the selection function. This is in part because broader selection functions grant a larger $w_{g\mu}(\theta)$ term which dominates at low redshifts. In the region near the baryon bump, this term is more strongly scale dependent than $w_{\mu\mu}(\theta)$. Additionally, the baryon peak is sharper and the amplitude of $w_{gg}(\theta)$ is larger for narrower selection functions (e.g. Figure \ref{th2zoomfig}), both of these factors make it more difficult for magnification to effect the peak location.% For example, at $z_0=3.0$ $w_{\mu\mu}(\theta,\sigma=0.07)\approx w_{\mu\mu}(\theta,\sigma=0.30)$ and at this redshift $w_{\mu\mu} >> 2w_{g\mu}$ for both values of $\sigma$. The lensing correction to $w_{gg}(\theta)$ is then nearly the same for both $\sigma=0.07$ and $\sigma=0.30$, however the fractional shift in peak height, location and width is much larger for $\sigma=0.30$ than for $\sigma=0.07$ since $w_{gg}(\theta,\sigma=0.30) < w_{gg}(\theta,\sigma=0.07)$. 

Magnification shifts $\theta_{BAO}$ by $<0.5\%$ for all $\sigma$ and $z_0$ values we considered.  The changes to the width and amplitude are much larger. We should point out that as the width of the selection function is increased the baryon bump is also broadened making the peak harder to identify -- this is true regardless of whether magnification bias is present. This is illustrated in Figure \ref{th2zoomfig}, where we show $\theta^2w(\theta)$ for a narrow angular range for each $\sigma$. One can see that for $\sigma=0.30$ there is no maximum at this redshift. 

%Despite the large changes seen in the angular power spectrum with $(5s-2)/b=1$ (seen in Figure \ref{CnnCnoBfig}), the changes in the peak location of the angular correlation function at these scales is relatively small - the dominant effect is to change the amplitude. It appears that changes to $\theta_{BAO}$ $\gsim 0.2\%$ are present only for extremely broad selection functions ($\sigma \gsim 0.30$) and/or galaxies with large slope to bias ratio $(5s-2)/b\sim 2$. 
%Figures \ref{th2Diff07}-\ref{th2Diff30} summarize the changes to the peak location and height for several redshifts. 

We have shown that the change in location of the baryon bump found in our calculated $w_{nn}(\theta)$ and $w_{gg}(\theta)$ is likely to be small. However, an actual measurement of $\theta_{BAO}$ will involve fitting a curve to data points with error bars. From this perspective, magnification bias -- which adds a scale and source population dependent correction to $w_{gg}(\theta)$ -- may be of more concern. This fact is illustrated in the lower panel of Figure \ref{wggNear}. Here we show the difference between the two-point functions with and without magnification ($w_{nn}-w_{gg}$). Although this difference is slowly varying in the region around the baryon peak, the value of the difference is not constant. In the regime very near the peak the offset between $w_{nn}(\theta)$ and $w_{gg}(\theta)$ can be approximated as a straight line. The precise slope of this line will depend on the population of galaxies via $b$ and $s$. Interestingly, the magnification correction to $\theta^2w(\theta)$ is closer to a constant than it is for $w(\theta)$. When the baryon bump in $\theta^3w(\theta)$ is considered we find the magnification correction is more strongly scale dependent than for $\theta^2w(\theta)$. Thus the changes to the baryon peak location in $\theta^3w(\theta)$ are significantly larger (by a factor of $5-6$). We emphasize then, that while Figure \ref{th2DiffAll} gives an indication of the effect of magnification bias on the baryon bump, the actual effect of magnification on measurements of the acoustic scale will depend on the method by which the peak location is fitted from the data.  When using the BAO peak as a standard ruler this scale and population dependent bias should be taken into account. 

\section{discussion}
\label{discussion}
The angular two-point correlation function and the angular power spectrum are important cosmological statistics, but a complete understanding of systematics is necessary to use these to derive precise constraints on cosmological parameters. Magnification bias, a gravitational lensing correction to the observed number density of sources, has been known to change the correlation function at high redshifts \cite{Villumsen1995,VFC97,MJV98,MJ98}. We have examined and quantified how magnification bias changes the shape of the angular correlation function and power spectrum.  The effect of magnification is to adjust both the scale-dependence and amplitude of the correlation function. We have shown that the scale-dependence changes the location of the matter-radiation equality peak in the angular power spectrum, and can potentially lead to a bias in determining the location of the baryon bump in the angular correlation function (see Figures \ref{CnnCnoBfig}, \ref{wggNear}, \ref{th2DiffAll} and Table \ref{lpeaktable}). Magnification bias becomes important at high redshifts ($z\gsim 1.5$) with the most drastic scale dependent changes occurring at low multipoles $\ell \lsim 100$. Precisely how, if ignored,  magnification would bias measurements of the matter-radiation equality scale or the BAO size will likely depend on how these quantities are extracted from the correlation function as well as the population of galaxies or quasars used for the measurement. Finally, the magnification terms are sensitive to non-linear evolution of structure at low redshifts even for sources at high redshift. Consequently, the presence of magnification bias changes the redshift and scale for which non-linear evolution of the power spectrum is visible.

Magnification bias is a source-population dependent effect, that is the magnitude (and even the sign) of the correction depend on the population of galaxies or quasars being observed. On the one hand, one may be able to select a population of sources with number count slope $s=2/5$, in which case magnification bias vanishes. However, such a selection may reduce the number of sources available for analysis so this is not necessarily the best option. Current measurements of the galaxy 
angular correlation are at redshifts $z<1$, so the correction from magnification to these observations is expected to be negligible. Projections for measurements of galaxy clustering from future high redshift galaxy surveys will need to address the effect of magnification bias. 

Finally, let us reiterate that this paper focuses exclusively on the angular correlation.
The effect of magnification bias on the 3D correlation has some surprising new features.
For instance, in certain situations, magnification bias can be important even for low
redshift measurements. This was originally noted by \cite{Matsubara} and is further analyzed in two separate papers \cite{3Dpaper1,3Dpaper2}.

\acknowledgments

LH thanks Ming-Chung Chu and the Institute of Theoretical Physics
at the Chinese University of Hong Kong
for hospitality where part of this work was done.
Research for this work is supported by the DOE, grant DE-FG02-92-ER40699,
and the Initiatives in Science and Engineering Program
at Columbia University.
EG acknowledges support from Spanish Ministerio de Ciencia y
Tecnologia (MEC), project AYA2006-06341 with EC-FEDER funding, and
research project 2005SGR00728 from  Generalitat de Catalunya.

\bigskip
\newcommand\spr[3]{{\it Physics Reports} {\bf #1}, #2 (#3)}
\newcommand\sapj[3]{ {\it Astrophys. J.} {\bf #1}, #2 (#3) }
\newcommand\sapjs[3]{ {\it Astrophys. J. Suppl.} {\bf #1}, #2 (#3) }
\newcommand\sprd[3]{ {\it Phys. Rev. D} {\bf #1}, #2 (#3) }
\newcommand\sprl[3]{ {\it Phys. Rev. Letters} {\bf #1}, #2 (#3) }
\newcommand\np[3]{ {\it Nucl.~Phys. B} {\bf #1}, #2 (#3) }
\newcommand\smnras[3]{{\it Monthly Notices of Royal
        Astronomical Society} {\bf #1}, #2 (#3)}
\newcommand\splb[3]{{\it Physics Letters} {\bf B#1}, #2 (#3)}

\newcommand\AaA{Astron. \& Astrophys.~}
\newcommand\apjs{Astrophys. J. Suppl.}
\newcommand\aj{Astron. J.}
\newcommand\mnras{Mon. Not. R. Astron. Soc.~}
\newcommand\apjl{Astrophys. J. Lett.~}
\newcommand\etal{{\it et al.}}


\begin{thebibliography}{99}
%about the power spectrum
\bibitem{Peebles1973} P. J. E. Peebles, \apj 185, 413 (1973); M. G. Hauser, P. J. E. Peebles \apj 185, 757 (1973); E. J. Groth, P. J. E. Peebles \apj 217, 385 (1977); M. Tegmark \prl 79, 3806 (1997); M. Tegmark, A. J. S. Hamilton, M. A. Strauss, M. S. Vogeley, A. S. Szalay \apj 499, 555 (1998); Y. Wang, D. N. Spergel, M. A. Strauss \apj 510, 20 (1999); W. Hu, D. J. Eisenstein, M. Tegmark, M. White \prd 59, 023512 (1998); D. J. Eisenstein, W. Hu, M. Tegmark, \apj 518, 2 (1999)
%\bibitem{PeebHauser1973} M. G. Hauser, P. J. E. Peebles \apj 185, 757 (1973)
%\bibitem{GrothPeeb1977} E. J. Groth, P. J. E. Peebles \apj 217, 385 (1977)
%\bibitem{Teg1997} M. Tegmark \prl 79, 3806 (1997)
%\bibitem{Teg1998} M. Tegmark, A. J. S. Hamilton, M. A. Strauss, M. S. Vogeley, A. S. Szalay \apj 499, 555 (1998)
%\bibitem{Wang1999} Y. Wang, D. N. Spergel, M. A. Strauss \apj 510, 20 (1999)
%\bibitem{HuEisen1999} W. Hu, D. J. Eisenstein, M. Tegmark, M. White \prd 59, 023512 (1999)
%\bibitem{EisHuTeg1999} D. J. Eisenstein, W. Hu, M. Tegmark, \apj 518, 2 (1999)
\bibitem{Cooray06} A. Cooray, \apj 651, L77 (2006)



\bibitem{BAO1} J. Silk, \apj 151, 459 (1968)
\bibitem{BAO2} P. J. E. Peebles, J. T. Yu, \apj 162, 815 (1970)
\bibitem{BAO3} R. A. Sunyeav, Ya. B. Zel'dovich, Astrophysics and Space Science, 7, 3S (1970)
\bibitem{BAO4} J. R.  Bond, G. Efstathiou, \apj 285, 45 (1984)
\bibitem{BAO5} J. A. Holtzman, \sapjs 71, 1 (1989)
\bibitem{EH98} Eisenstein, D. J., Hu, W., \apj  496, 605 (1998)
\bibitem{BAO6} A. Meiksin, M. White, J. A. Peacock \mnras 304, 851 (1999)

\bibitem{WMAPI} D. Spergel \etal \apjs 148, 175 (2003)
\bibitem{WMAPIII} D. Spergel \etal, \apj, in press



 \bibitem{BAOth1} D. J. Eisenstein, W. Hu, M. Tegmark \apj 504, 57 (1998); H.- J. Seo, D. J. Eisenstein \apj 598, 720 (2003); E. V. Linder, \prd 68, 083504 (2003); T. Matsubara, A. S. Szalay, \prl 90, 021302 (2003); C. Blake, K. Glazebrook \apj 594, 655 (2003); W. Hu, Z. Haiman, \prd 68, 063004 (2003); T. Matsubara \apj 615, 573 (2004);  H.- J. Seo, D. J. Eisenstein \apj 633, 575 (2005); K. Glazebrook, C. Blake \apj 631, 1 (2005); M. White, Astropart. Phys., 24, 334 (2005); C. Blake, S. Bridle, \mnras 363, 1329 (2005); R. Angulo, C.M. Baugh, C.S. Frenk, R. G. Bower, A. Jenkins, S. L. Morris \mnras 262, L25 (2005); C. Blake, D. Parkinson, B. Basset, K. Glazebrook, M. Kunz, R. C. Nichol \mnras 365, 255 (2006); D. Dolney, B. Jain, M. Takada \mnras 366, 884 (2006);  D. Dolney, B. Jain, M. Takada \mnras 366, 884 (2006); D. Jeong, E. Komatsu, \apj 651, 619 (2006); E. Huff, A. E. Schulz, M. White, D. Schlegel, M. S. Warren, Astropart. Phys. 26, 351 (2007); R. Angulo, C. M. Baugh, C. S. Frenk, C. G. Lacey \mnras in press (2007)
% \bibitem{BAOth2} H.- J. Seo, D. J. Eisenstein \apj 598, 720 (2003)
 %\bibitem{BAOth3} E. V. Linder, \prd 68, 083504 (2003)
 %\bibitem{BAOth4} T. Matsubara, A. S. Szalay, \prl 90, 021302 (2003)
 %\bibitem{BAOth5} C. Blake, K. Glazebrook \apj 594, 655 (2003)
 %\bibitem{BAOth6} W. Hu, Z. Haiman, \prd 68, 063004 (2003)
 %\bibitem{BAoth7} T. Matsubara \apj 615, 573 (2004)
 %\bibitem{BAOth8}  H.- J. Seo, D. J. Eisenstein \apj 633, 575 (2005)
 %\bibitem{BAOth9} K. Glazebrook, C. Blake \apj 631, 1 (2005)
%\bibitem{BAOth10} M. White, Astropart. Phys., 24, 334 (2005)
 %\bibitem{BAOth11} C. Blake, S. Bridle, \mnras 363, 1329 (2005)
 %\bibitem{BAOth12} R. Angulo, C.M. Baugh, C.S. Frenk, R. G. Bower, A. Jenkins, S. L. Morris \mnras 262, L25 (2005)
 %\bibitem{BAOth13} C. Blake, D. Parkinson, B. Basset, K. Glazebrook, M. Kunz, R. C. Nichol \mnras 365, 255 (2006)
 %\bibitem{BAOth14} D. Dolney, B. Jain, M. Takada \mnras 366, 884 (2006)
 %\bibitem{BAOth15} D. Jeong, E. Komatsu, \apj 651, 619 (2006)
 %\bibitem{BAOth16} E. Huff, A. E. Schulz, M. White, D. Schlegel, M. S. Warren, Astropart. Phys. 26, 351 (2007)
 %\bibitem{BAO17}  R. Angulo, C. M. Baugh, C. S. Frenk, C. G. Lacey \mnras in press (2007)

 %BAO detection
  \bibitem{BAOmeas1} S. Cole, \etal, \mnras 362, 505 (2005)
 \bibitem{BAOmeas2} D. J. Eisenstein, \etal  \apj 633,560 (2005)
 \bibitem{BAOmeas3} G. Huetsi, \AaA submitted (2005) astro-ph$/0507678$
 \bibitem{BAOmeas4} N. Padmanabhan, \etal \mnras submitted (2006) astro-ph$/0605302$ 
 \bibitem{BAOmeas5} W. Percival, \etal \apj 657, 51 (2007)
%magnification bias predictions
\bibitem{Gunn67} J. E. Gunn, \apj 147 61 (1967)
\bibitem{Narayan1989} E. L. Turner, J. P. Ostriker, J. R. Gott, \apj 284, 1 (1984);
R. L. Webster, P. C. Hewett, M. E. Harding, G. A. Wegner, Nature  336, 358 (1988);
W. Fugmann, \AaA 204, 73 (1988);
R. Narayan, \apjl 339, 53 (1989); P. Schneider, \AaA 221, 221 (1989).
\bibitem{BTP1996} T. J. Broadhurst, A. N. Taylor, J. A. Peacock, \apj 438, 49 (1996)

\bibitem{Villumsen1995} J. V. Villumsen, unpublished preprint (1995) astro-ph$/9512001$
\bibitem{VFC97} J. Villumsen, W. Freudling, L. N. da Costa, \apj 481, 578 (1997)
\bibitem{MJV98} R. Moessner, B. Jain, J. Villumsen,  \mnras 294, 291 (1998)
\bibitem{MJ98} R. Moessner, B. Jain,\mnras 294, L18 (1998)
%galaxy quasar
\bibitem{EGmag03} E. Gazta\~{n}aga,  \apj 589, 82 (2003)
\bibitem{ScrantonSDSS05} R. Scranton, \etal, \apj 633, 589 (2005)
\bibitem{Myers2003} A. D. Myers, \etal \mnras 342, 467 (2003)
\bibitem{menard} B. Menard, M. Bartelmann, \AaA 386, 784 (2002).
\bibitem{JSS03} B. Jain, R. Scranton, R. V. Sheth, \mnras 345, 62  (2003) 
\bibitem{Matsubara} T. Matsubara, \apj 537, L77 (2000)
\bibitem{3Dpaper1} L. Hui, E. Gaztanaga, M. LoVerde, submitted to \prd (2007) arxiv$/0706.1071$
\bibitem{3Dpaper2} L. Hui, E. Gaztanaga, M. LoVerde,  submitted to \prd (2007b) arxiv$/0710.419$
\bibitem{otherpaper} A. Vallinotto, S. Dodelson, C. Schimd, J. P. Uzan  (2007) astro-ph$/0702606$
\bibitem{dodelson} For a review, see S. Dodelson, Modern Cosmology, Academic Press (2003).
\bibitem{RS3} R. E. Smith, R. Scoccimarro, R. K. Sheth, \prd 75, 063512 (2007) 
\bibitem{Limber} D. N. Limber, \apj 119, 655 (1954)
% transfer function and non-linear power spectrum

\bibitem{BBKS} J. M. Bardeen, J. R. Bond, N. Kaiser, A. S. Szalay, \apj 304, 15B (1986)
\bibitem{GammaSugiyama} N. Sugiyama, \sapjs 100, 281 (1995)
\bibitem{Smithetal} R. E. Smith, \etal \mnras 341 1311  (2003)
\bibitem{mepaper} M. LoVerde, L. Hui, E. Gazta\~{n}aga, \prd 75, 043519 (2007)

%\bibitem{NumericalRecipes} W. H. Press, S. A. Teukolsky, W. T. Vetterling, B. P. Flannery, Numerical Recipes in FORTRAN: The Art of Scientific Computing, (Cambridge: University Press, Ñc1992, 2nd ed.) 

\end{thebibliography}
\end{document}